\documentclass[useAMS,usenatbib]{mnras}
\usepackage{amsmath}
\usepackage{graphicx,txfonts,fleqn,enumerate,color,bm}
\usepackage{color}
\usepackage{ulem}

\newcommand*  {\twovector}[2] {{\begin{pmatrix} $1 \\ $2 \end{pmatrix}}}

\renewcommand {\emph}[1]  {\textit{#1}}

\title[Should $N$-body integrators be symplectic everywhere in phase space?]
{Should $N$-body integrators be symplectic everywhere in phase space?}

\author[David M. Hernandez]
	{David M. Hernandez \thanks{Email: dmhernandez@cfa.harvard.edu}  \\
	Harvard--Smithsonian Center for Astrophysics, 60 Garden St., MS 51, Cambridge, MA 02138, USA \\
	}
\begin{document}

\maketitle

\label{first page}
\begin{abstract}
Symplectic integrators are the preferred method of solving conservative $N$-body problems in cosmological, stellar cluster, and planetary system simulations because of their superior error properties and ability to compute orbital stability.  Newtonian gravity is scale free, and there is no preferred time or length scale: this is at odds with construction of traditional symplectic integrators, in which there is an explicit timescale in the time-step.  Additional timescales have been incorporated into symplectic integration using various techniques, such as hybrid methods and potential decompositions in planetary astrophysics, integrator sub-cycling in cosmology, and block time-stepping in stellar astrophysics, at the cost of breaking or potentially breaking symplecticity at a few points in phase space.  The justification provided, if any, for this procedure is that these trouble points where the symplectic structure is broken should be rarely or never encountered in practice.  We consider the case of hybrid integrators, which are used ubiquitously in astrophysics and other fields, to show that symplecticity breaks at a few points  are sufficient to destroy beneficial properties of symplectic integrators, which is at odds with some statements in the literature.  We show how to solve this problem in the case of hybrid integrators by requiring Lipschitz continuity of the equations of motion.  For other techniques, like time step subdivision, {consequences to this this problem are not explored here}, and the fact that symplectic structure is broken should be taken into account by $N$-body simulators, who may find an alternative non-symplectic integrator performs similarly.
\end{abstract}
\section{Introduction}
The $N$-body problem is the problem of solving for the motion in time of $N$-point particles, interacting through pairwise gravitational forces.  This problem describes, at 0th order, a wide range of astrophysical dynamics, from the motion of stars in galaxies, to planets orbiting stars \citep{HH03}.  Since the $N$-body problem is {generally} non{-}integrable, astrophysicists have employed a number of approximate techniques to solve it.  Perhaps the most robust and successful program to solve the $N$-body problem is to use numerical integrators which approximately solve the system of $6N$ $N$-body ordinary differential equations.  Other approaches exist and have been used with varied levels of success.  One can solve the collisionless Boltzmann equation for dark matter \citep{HAK13}, treating the dark matter as a collisionless fluid.  The Fokker-Planck approximation solves the Boltzmann equation with a collision term and can be applied to stellar clusters \citep{BT08}.  $N$-body secular behavior can be studied by orbit-averaging Hamiltonians \citep{HP16}.  

Numerical integrators suffer from errors due to their approximations, and also from errors due to the limitations of computer memory and the amount of digits a computer can store for a number.  Thus, their arithmetic is not exactly precise.  Symplectic integrators have been employed in astrophysics since at least the 1980's and have been a preferred integrator for describing a wide range of dynamics.  Symplectic integrators preserve geometric features \citep{hair06} in phase space of the ordinary differential equations they solve.  As a result, they have excellent error properties and reproduce orbits faithfully.  Symplectic integrators are used for long term investigations of Solar System chaos and for cosmological simulations, among many other uses.  In their traditional construction, for example, by operator splitting \citep{Y90}, they require an input timescale in the form of a timestep.  However, Newtonian gravity has no characteristic time of length scale so this input timestep is generally unphysical.  This problem can be mitigated by decomposing the $N$-body problem into $2$-body problems \citep{HB15,H16}.  For problems with well-defined timescales, like the planets orbiting the Sun, a traditional symplectic integrator with a timestep of, say, 1 year, is highly successful for studying the long-term dynamics of this problem \citep{WH91,HD17}.  For other problems that involve binary stars or scattering of planets, a given timestep runs the risk of being unable to resolve these dynamical phenomena.  At the same time, using too small a timestep can be computationally prohibitive.  

Because of the extreme limitation of specifying a timescale, astrophysicists developed new methods that can introduce additional timescales into symplectic integrators.  Block timestep integrators \citep{farr07}, hybrid symplectic integrators \citep{C99}, timestep subdivision methods \citep{S05}, and potential decomposition methods \citep{DLL98} are examples of methods that have become standard in the dynamical astrophysics toolbox.  However, there is a cost to using these methods: symplecticity can be broken at some number of phase space points.  Some have justified the use of these methods by noting the problem points are rare and should not pose practical problems, while others have not acknowledged any potential problems.

We test the assumption that breaking symplecticity at a finite number of points is not problematic in this paper.  While the results apply to all the multiple timescale methods above, we focus on hybrid integrators in this paper.  These integrators use a transition function to switch from a  method with a long time scale to a method that resolves short timescales.  The transition function determines if the integrator breaks symplecticity at a number of points, while remaining symplectic elsewhere.  We show the hybrid integrators with symplecticity breaks perform substantially worse than the fully symplectic hybrid integrators.  $N$-body code developers and users should be aware of the impact of breaking symplecticity at some points which can make their supposed symplectic code behave similarly to non-symplectic alternatives.  For hybrid integrators, fortunately, full symplecticity is ensured by enforcing Lipschitz continuity of the equations of motion.  In the case of block or multiple time-stepping schemes, we do not present a solution to their limitation.

In Section \ref{sec:Kepler} we discuss the Kepler problem and how symplectic Euler can be used to solve it.  We define symplectic maps according to whether the Hamiltonian ordinary differential equations are Lipschitz continuous.  Then we present a hybrid integrator to solve the Kepler problem.  In Section \ref{sec:sfunc} we present the switching functions we use, some yielding symplectic integrators and some not.  In Section \ref{sec:numex}, we present numerical experiments with the hybrid integrators.  We show only the fully symplectic ones have the desired stability properties.  We conclude in Section \ref{sec:conc}.

\section{Solving the Kepler problem}
\label{sec:Kepler}
The Kepler problem describes the two-body problem, in which the bodies are treated as point particles, and they interact through Newtonian gravity.  This problem has six degrees of freedom, three describing the relative motion of the bodies, and three describing the center of mass coordinates.  The {latter} three are removed by a Galilean transformation to the center of mass frame.  In spherical polar coordinates, we use the fact {that} the $z$ angular momentum is conserved to deduce the motion is planar and rotate to the plane of motion.  In polar coordinates, the angular momentum is seen to be a constant, so our final system is an integrable, one-degree of freedom system.  We choose a simplified system of units, similar to $N$-body units \citep{HH03}{.  In that case,} the gravitational constant $G=1$, the total mass $M=1$ and total energy is $E = -1/4$ (putting a constraint on the virial radius).  In our case, the reduced mass, $\mu = 1$, the gravitational constant is reciprocal to the total mass $GM = 1$, and the semi-major axis $a = 1$.  The Hamiltonian is then,
\begin{equation}
H_{\mathrm{Kep}} = \frac{p^2}{2} + \frac{L^2}{2r^2} - \frac{1}{r},
\label{eq:hamilt}
\end{equation}
where $r$ is the distance from the focus at the origin, and $p$ is its conjugate momentum, $p = \mu \dot{r}$, where $\dot{r}$ indicates the time derivative of $r$.  $L$ is the angular momentum{, conjugate to the polar angle.}  In terms of the eccentricity, it is  $L = \sqrt{1 - e^2}$ for elliptic motion, and $L = \sqrt{e^2 - 1}$ for hyperbolic motion.  For parabolic motion, $L = 0$.  {Hyperbolic orbits have $H_{\text{Kep}} = +1/2$.}  We concern ourselves with elliptic motion in this work to study periodic motion.  The value of the Hamiltonian is {$H_{\text{Kep}} = -1/2$} and the period is $P = 2 \pi$.  

In this paper, we will also relax the assumption that $L$ is constant.  In this case, the Hamiltonian has two degrees of freedom, described by the vector $(x_1,x_2,p_1,p_2)$.  In the same units, the Hamiltonian is
\begin{equation}
H_{\mathrm{Kep2}} = \frac{p_1^2 + p_2^2}{2} - \frac{1}{\sqrt{x_1^2 + x_2^2}}.
\label{eq:kep2}
\end{equation}
The coordinate systems are related by $r = \sqrt{x_1^2 + x_2^2}$ and $p = (x_1 p_1 + x_2 p_2)/r$.

\subsection{Arnold--Liouville theorem and symplecticity}
\label{sec:sympint}

A Hamiltonian is a function $H(p,q,t)$ that does not need to be continuous.  Hamilton's equations are a coupled set of first order ordinary differential equations (ODEs): 
\begin{equation}
\begin{aligned}
\dot{q}_i &=  \frac{\partial H}{\partial p_i}, \quad \text{and} \\
\dot{p}_i & = -  \frac{\partial H}{\partial q_i}.
\label{eq:hamilteq}
\end{aligned}
\end{equation}
Consider the Hamiltonian, $H = \theta(q) + p$, where $\theta$ is the Heaviside function.  In the motion in time, there is a jump discontinuity in $q$, so that the trajectory is defined everywhere except at a point $q_0$.  In astrophysics, we are frequently concerned with periodic orbits for which action-angle variables are defined.  Two examples of periodic orbits are Kepler orbits and the orbits in St\"{a}ckel potentials, which are models of galactic potentials.  The notion of an integrable Hamiltonian exists if the Arnold--Liouville theorem holds, which places restrictions on the Hamiltonian.  \cite{AX16} show it is sufficient that the Hamiltonian is of class $C^{1,1}$ for the Arnold--Liouville theorem to hold.  $C^{1,1}$ means the Hamiltonian is at least continuously differentiable in phase space once, or $C^1$, and those derivatives are Lipschitz continuous.  This is weaker than a $C^2$ requirement of the Hamiltonian.  Lipschitz continuity is a stronger condition than continuity.  It places a bound on the variation of a function and ensures the existence and uniqueness of a solution to ODEs like Eq. \eqref{eq:hamilteq}, according to the Cauchy--Lipschitz theorem.  

Let $f(x,t)$ be one of the $2n$ differential equations in \eqref{eq:hamilteq}, where $x$ is a phase space vector. {The number of degrees of freedom is $n$}.  If $f$ is Lipschitz continuous, there exists a positive constant $M$ such that,
\begin{equation}
| f(x_1,t) - f(x_2,t) | \le M \times\text{norm}(x_1 - x_2),
\label{eq:lips}
\end{equation}
for all $x_1,x_2$.  $\text{norm}(y)$ is the maximum of $y$.  In the  Lipschitz condition, the independent variable, time in our case, is less of a concern, and we will concern ourselves with autonomous Hamiltonians anyway.  The Kepler Hamiltonian is $C^\infty$ when $r$ is restricted to $r > 0$.  Similarly, the $N$-body Hamiltonian is $C^\infty$ for the domain where particle separations are greater than 0.

Symplecticity is a statement about conservation of invariants; when they are smooth, they are denoted Poincar\'{e} invariants.  Hamiltonian flow is symplectic.  In classical mechanics and astrophysics, the symplecticity is often described by the Jacobian matrix \citep{farr07,SW01,HB15,hair06}.  If a Hamiltonian is smooth enough, any map of phase space onto itself has an associated Jacobian matrix $J$ of size $2n \times 2n$.  This Jacobian matrix has $2n^2 + n$ constraints if the map is derived from a time-independent Hamiltonian.  The constraints are summarized by,
\begin{equation}
    J^{\dagger} \Omega J = \Omega,
    \label{eq:sympmat}
\end{equation}
where the form of the constant matrix $\Omega$ depends on the phase space basis.  By conserving Poincar\'{e} invariants, symplectic integrators can aid the study of orbital stability while conventional integrators cannot, and symplectic integrators ensure bounded energy error \citep{hair06}.  

We find our first difficulty: if the Hamiltonian ODEs are Lipschitz continuous, Rademacher's theorem guarantees the existence of a Jacobian matrix almost everywhere, but not everywhere.  This is a limitation of using Eq. \eqref{eq:sympmat} as a description of symplecticity.  In fact, a notion of symplecticity exists even when the Hamiltonian is $C^0$ \citep{BHS18}.  In this paper, we define symplectic integrators as those associated with Hamiltonians of smoothness of at least $C^{1,1}$, because these are known to satisfy the Arnold--Liouville theorem, which applies to the $N$-body problem.  Our numerical experiments support that this minimum smoothness is required for periodic orbits to exist, with some caveats we describe.  

\cite{HW83} studied a map describing a billiard table with discontinuous curvature.  The Jacobian was not defined everywhere and the emergence of chaos was observed.
\subsection{Symplectic integration of Kepler problem}
Eq. \eqref{eq:hamilt} is solved with Hamilton's equations:  

\begin{equation}
\begin{aligned}
\dot{r} &= p, \\
\dot{p} & = - \frac{1}{r^2} + \frac{L^2}{r^3}.
\label{eq:eom}
\end{aligned}
\end{equation}
Solving for the motion in time explicitly is possible, since the problem is integrable, but is inconvenient.  A more convenient solution involves solving the implicit Kepler equation, whose form depends on whether the motion is elliptic or hyperbolic.  The Kepler equation is implicit, so to solve it, one usually guesses a solution, and iterates to refine it.  By transforming the time variable to a universal variable \citep{dan88}, a generalization of the Kepler equation can be written that doesn't depend on whether the motion is bound.  

For the purposes of exploring hybrid symplectic integrators, we instead solve \eqref{eq:hamilt} using a symplectic Euler method \citep{hair06}, using operator splitting.  {Thus, we} now solve two Hamiltonians,
\begin{equation}
H_1 = \frac{p^2}{2} \quad \text{and} \quad H_2 = \frac{L^2}{2r^2} - \frac{1}{r},
\label{eq:simpsplit}
\end{equation}
successively, which gives an approximation to the solution \eqref{eq:hamilt}; the energy error is accurate to first order in the time of integration, or the timestep \citep{hair06}.  Note $H_{\mathrm{Kep}} = H_1 + H_2$.  Of course, $H_1$ and $H_2$ are still integrable, but their respective solutions from Hamilton's equations are now trivial to write.  We require $H_1$ and $H_2$ to be Lipschitz in order for the integrator to be considered symplectic according to our definition.  For an example of how symplectic Euler works, see \cite{HB15}.  In more precise notation, let the phase space be $z$, $z^\prime$ the updated phase space, $h$ the timestep, and {$N = t/h$}, where $t$ is the total time, a multiple of $h$.  The integrator can be represented as,
\begin{equation}
z^\prime = \left(e^{h \hat{H}_1} \circ e^{h \hat{H}_2} \circ \right)^N z,
\label{eq:phspup}
\end{equation}
where $\hat{a}$ is an operator.  A symplectic map can be studied outside the context of a computer and its finite memory.  Its dynamics are specified exactly through its modified differential equation \citep{HB18,hair06}.
\subsection{Hybrid symplectic integration of the Kepler problem}
Now we do a hybrid \citep{C99,K00,DLL98,H16,Wisdom2017} symplectic integration of the Kepler problem.  Instead of splitting $H_{\mathrm{Kep}}$ into \eqref{eq:simpsplit}, we can instead split into
\begin{equation}
H_1 = \frac{p^2}{2} - \frac{K(r)}{r} \quad \text{and} \quad H_2 = \frac{L^2}{2r^2} - \frac{1}{r}\left(1 - K(r) \right).
\label{eq:split1}
\end{equation}
(If $K$ does not satisfy certain smoothness requirements, the integrator will not be symplectic according to our definition).  The hybrid integrator transfers pieces from $H_1$ to $H_2$ and vice-versa.  In real world applications, this strategy allows one to resolve small timescales in $N$-body problems when needed, but not all that time so that the integrator is too costly and impractical.  The range of $K$ is $[0,1]$.   

The solution from the $H_2$ equations of motion {is} easy to write:    
\begin{equation}
\begin{aligned}
r^\prime &= r, \\
p^\prime &= p + h\left( \frac{L^2}{r^3} - \frac{1}{r^2}(1 - K) - \frac{dK/dr}{r} \right).
\end{aligned}
\end{equation}
The solution from $H_1$ is easy to write when $K = dK/dr = 0$:
\begin{equation}
\begin{aligned}
r' &= r + h p, \\
p' &= p.
\label{eq:simpmap}
\end{aligned}
\end{equation} 
But suppose $K$ becomes non-zero during the step; then \eqref{eq:simpmap} will no longer be valid; the correct equations to solve will be,
\begin{subequations}
\begin{flalign}
\dot{r} &= p, & \\
\dot{p} & = \frac{dK/dr}{r} - \frac{K}{r^2} , &
\label{eq:compmap2}
\end{flalign}
\label{eq:compmap}
\end{subequations}
with a more complicated solution map in general.  The map is guaranteed to exist, however, because the Hamiltonian is integrable.  Practitioners would want to use map \eqref{eq:simpmap} as much as possible for fast calculations, but switch to map \eqref{eq:compmap} otherwise, which is always correct. However, a failsafe way to guarantee {that} we haven't used map \eqref{eq:simpmap} incorrectly does not exist \citep{C99,Wisdom2017}.  We will not focus on this limitation of hybrid symplectic integrators here, but instead overcome this limitation by always solving \eqref{eq:compmap}, which is inefficient.

We can solve \eqref{eq:compmap} using a high accuracy method like Bulirsch-Stoer, which uses Richardson extrapolation to estimate a map in the limit of the stepsize going to 0.  We will also experiment later with adaptive Runge--Kutta--Fehlberg methods, which combines a fourth and fifth order Runge--Kutta method and is more suitable for non-smooth functions \citep{press02}, and other alternatives.

\section{Switching functions}
\label{sec:sfunc}
\begin{figure}
    \centering
    \resizebox{0.99\columnwidth}{!}{\includegraphics{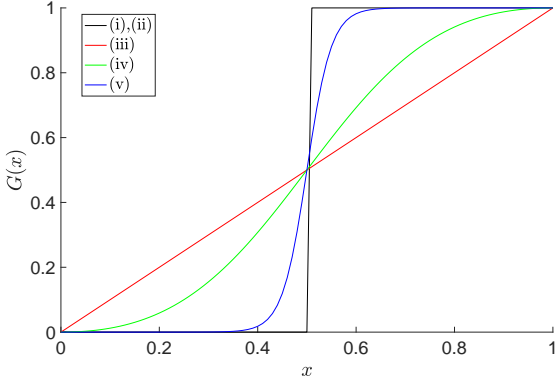}}
    \caption{Switching functions for the Kepler hybrid integrator Eq. \eqref{eq:split1}.  The roman numerals refer to the functions in the paper, \ref{item:heavi}, \ref{item:modheavi}, \ref{item:lin}, \ref{item:polyn}, and \ref{item:smooth}.  The functions \ref{item:heavi}, \ref{item:modheavi}, \ref{item:lin} do not give symplectic integrators, while functions \ref{item:polyn}, and \ref{item:smooth} do give symplectic algorithms.}
    \label{fig:fcts}
\end{figure}

We have yet to specify the form of $K(r)$, subject to the Hamiltonian constraints of Sec. \ref{sec:sympint}.  First, consider $x = r-1$.  In the exact problem described by $H_{\mathrm{Kep}}$, $x$ remains within the interval $(-e,e)$.  Let {$K(r) = G(r-1)$}. We consider the following $G(x)$:
\begin{enumerate}
\item A Heaviside function, $G(x) =  \theta(x - 0.5)$. If $x < 0.5$, $G(x) = 0$, and if $x \ge 0.5$, $G(x) = 1$.  $dG/dr = 0$ everywhere except at $x = 0.5$.  With this $G(x)$, Eq. \eqref{eq:compmap2} is not Lipschitz continuous.  Thus, the hybrid integrator \eqref{eq:split1} is nonsymplectic.  
\label{item:heavi}
\item A modified version of \ref{item:heavi}.  The difference with \ref{item:heavi} is that whatever value $G(x)$ takes at the start of the step remains the same during the step.  This choice cannot be described by Hamiltonians.
\label{item:modheavi}
\item A linear function.  If $x < 0$, $G(x) = 0$ and if $x > 1$, $G(x) = 1$.  Otherwise $G(x) = x$.  The $dK/dr$ term of \eqref{eq:compmap2} is not Lipschitz continuous so the hybrid integrator is not symplectic.  
\label{item:lin}
\item A polynomial function.  If $x < 0$, $G(x) = 0$ and if $x > 1$, $G(x) = 1$.  Otherwise $G(x) = {x^2}/({2x^2 - 2x + 1})$.  The term $-K/r^2$ of \eqref{eq:compmap2} is differentiable.  The term $f = (dK/dr)/r$ of \eqref{eq:compmap2} is continuous but not differentiable at $x = 0$ and $x = 1$, so we check its Lipschitz continuity.  For $0 < x < 1$,  
\begin{equation}
f =  \frac{2x}{(2x^2 - 2 x + 1)(x + 1))} - \frac{x^2(4x - 2)}{(2x^2 - 2x + 1)^2(x + 1)},
\end{equation}
which is $2x$ near $x = 0$.  For $x<0$, $f = 0$.  To verify Lipschitz continuity \eqref{eq:lips} near $x = 0$ , consider $x_1$ and $x_2$ near 0.  We need to find an $M$ such that 
\begin{equation}
|a x_2 - b x_1| \le  M |x_2  - x_1 |,
\label{eq:lippoly} 
\end{equation}
where $a$ and $b$ are $0$ or $2$ depending on whether $x_1,x_2$ are positive or negative.  If $a = b = 0$, then any $M$ satisfies \eqref{eq:lippoly}.  Otherwise $M \ge 2$ satisfies \eqref{eq:lippoly}.  Using similar, analysis, \eqref{eq:compmap2} is Lipschitz continuous near $x = 1$.   Thus, the integrator is symplectic with this switching function.     
\label{item:polyn}
\item A smooth $C^\infty$ function.   $G(x) = \frac{1}{2}(1+ \tanh(k(x- 0.5)))$ with $k = 5$.  $G(x)$ never actually reaches $0$ or $1$.   \eqref{eq:compmap} is infinitely differentiable.  This yields a symplectic integrator.  The larger $k$ is, the closer we approach function \ref{item:heavi} (while retaining smoothness).
\label{item:smooth}

\end{enumerate}
We plot the $G(x)$ in Fig. \ref{fig:fcts}.

\section{Numerical experiments}
\label{sec:numex}
We have defined five transition functions; two are not symplectic at just two points in phase space, two are symplectic everywhere, and \ref{item:modheavi} is not described by Hamiltonians.  Our goal is to study the error properties of the hybrid integrator \eqref{eq:split1}, over a long time in a periodic Kepler problem. We will set up our tests so that during an orbit, the integrator has to integrate over (but not necessarily exactly hit) one of the problem points twice.  Note that landing on the point itself is unlikely, but the chances are not 0.  The reason is that there are only a finite number of double precision numbers; this is a reason it would not be useful to characterize these discontinuities as having Lebesgue measure 0.  

Choose $e = 0.7$, $h = P/100$, and $t = 10 P$.  We initialize at apoapse: $r = 1 + e$ and $p = 0$.  Recall $r$ is in units of the semimajor axis.  We check during each time step deviations from symplecticity using Eq. \eqref{eq:sympmat}, but recall it is not a perfect measure of symplecticity, according to the discussion of Section \ref{sec:sympint}.  Eq. \eqref{eq:sympmat} describes $2 \times 2$ matrices for Hamiltonian \eqref{eq:hamilt}, with only one constraint, $||J|| = 1$.  To calculate deviations from this constraint, we consider four initial conditions, separated by a small distance in phase space from the phase space point at the start of the step.  The distance $\delta$ {cannot} be too small or the computer finite precision {will not} keep track of the differences in the trajectories.  If $\delta$ is too large, the trajectories are no longer nearby.  A balance must be struck between these two effects \citep[Section 5.7]{press02}, and an analytic answer to the optimal $\delta$ is difficult to obtain, so we search it numerically.  We use a symmetric and second order approximation to the Jacobian.  For one variable $z$, the approximation is:
\begin{equation}
    \frac{d z^\prime}{d z} = \frac{-\frac{1}{2} z^\prime(z - \delta) + \frac{1}{2} z^\prime(z + \delta)}{\delta} + \mathcal O(\delta^2),
    \label{eq:finitediff}
\end{equation}
where $z^\prime(z \pm \delta)$ means the initial condition to calculate $z^\prime$ was $z \pm \delta$. The $\delta^2$ indicates the approximation is second order.  For two variables, such as our Kepler case, there are four Jacobian elements:
\begin{equation}
\label{eq:jacdef}
J_{22} = \frac{\partial r^\prime}{\partial r},~~~~J_{11} = \frac{\partial p^\prime}{\partial p},~~~~J_{12} = \frac{\partial p^\prime}{\partial r},~~~~J_{21} = \frac{\partial r^\prime}{\partial p}.
\end{equation}

\begin{figure}
    \centering
    \resizebox{0.99\columnwidth}{!}{\includegraphics{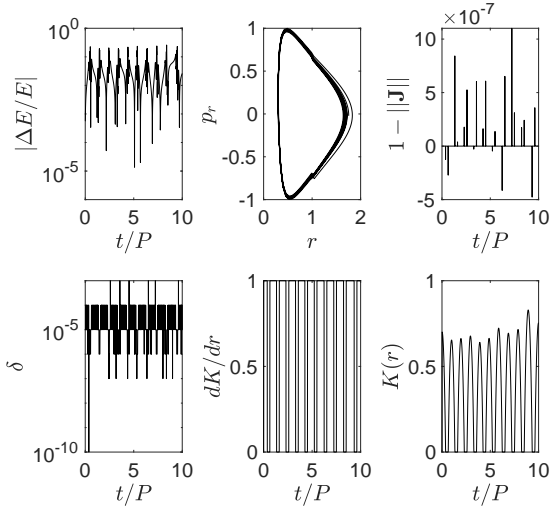}}
    \caption{Hybrid integrator test with transition function \ref{item:lin}.  This is not a symplectic integrator.  A Kepler problem is integrated.  The eccentricity is $e = 0.7$, the timestep is $h = P/100$, and the total time is $t = 10 P$, where $P = 2 \pi$, is the approximate period.  The panels show, from top left, going clockwise, the energy error in time, a phase space plot of the trajectory, the symplecticity error, $\delta$ used in the calculation of finite differences, the derivative of the transition function, and the transition function.  The energy error and phase space trajectory are not stable, as expected.  Apparent symplecticity breaks are found during integration over the problem point $x = 0$.}
    \label{fig:d0f1}
\end{figure}

\begin{figure}
    \centering
    \resizebox{0.99\columnwidth}{!}{\includegraphics{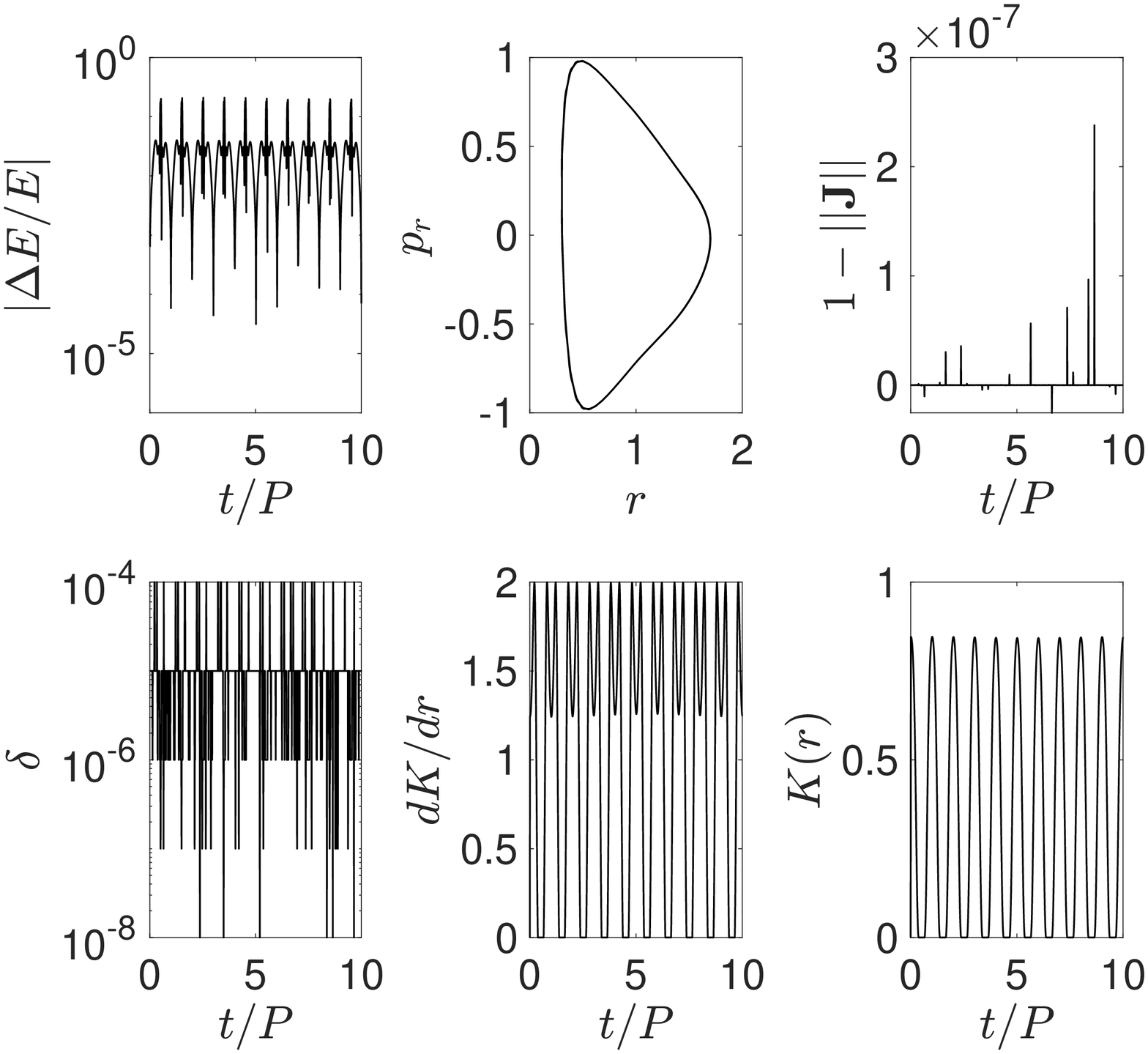}}
    \caption{{Same as in Fig. \ref{fig:d0f1}, but for an integration with transition function \ref{item:polyn}}.  This is a symplectic integrator.  The energy error and phase space trajectory are stable, as expected.  Symplecticity breaks appear to occur near the point $x = 0$ in phase space, even though the method is symplectic.}
    \label{fig:d1f1}
\end{figure}

\begin{figure}
    \centering
    \resizebox{0.99\columnwidth}{!}{\includegraphics{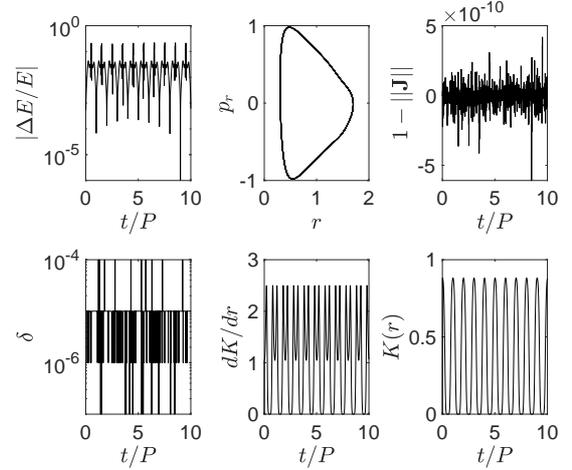}}
    \caption{{Same as in Fig. \ref{fig:d0f1}, but for an integration with transition function \ref{item:smooth}}.  This is a symplectic integrator.  The energy error and phase space trajectory are stable, as expected.  Symplecticity breaks appear to occur near the point $x = 0$ in phase space, even though the method is symplectic.}
    \label{fig:dallf1}
\end{figure}

We choose $\delta$ numerically by varying it in the range $(10^{-10},10^{-2})$ and choosing the value which gives the smallest $|R| = |1 - || J || |$ to ensure we're not overestimating $|R|$.  We calculate the energy error, $R$, $\delta$, $K$, and $dK/dr$ as a function of time.  Also we calculate the phase space trajectories of the orbits.  We show the results for smoothing functions \ref{item:lin}, \ref{item:polyn}, and \ref{item:smooth} in Fig's. \ref{fig:d0f1}, \ref{fig:d1f1}, and \ref{fig:dallf1}, respectively.  

\begin{figure}
    \centering
    \resizebox{0.99\columnwidth}{!}{\includegraphics{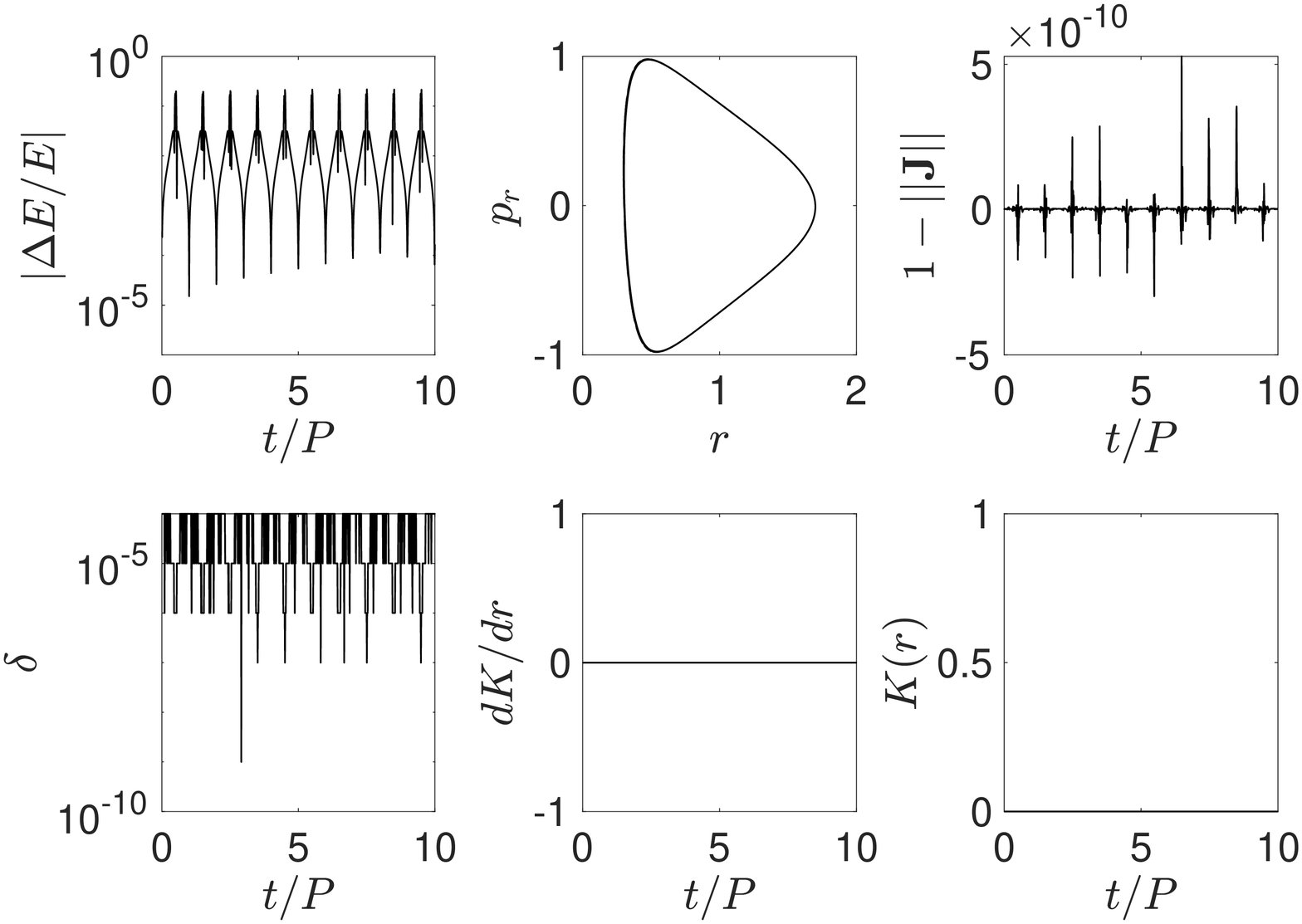}}
    \caption{{Same as in Fig. \ref{fig:d0f1}, but for a symplectic Euler integration.  We let} $K = dK/dr = 0$ in this case.  The energy error and phase space trajectory are stable.  10 apparent symplecticity breaks are found, {despite the method being symplectic.  Their} origin is different from those of the other figures.}
    \label{fig:euler}
\end{figure}

As expected, $K$ never reaches $1$, because $x = r-1$ does not exceed $0.7$ in the exact solution but does reach $0$.  The slope of $K$ can only be $1$ or $0$ for function \ref{item:lin}.  The energy error and phase space trajectory are stable in case \ref{item:polyn} and \ref{item:smooth}, because they are calculated using symplectic algorithms. We also see that the optimal $\delta$ varies across the spectrum of allowed values.  A more interesting story is shown in the plots of $R = 1 - ||J|| $.  There are 20 symplecticity error spikes, two per orbit, in all three cases, although the magnitude of those spikes decreases as the function becomes smoother.  The spikes happen during the transition from $K > 0$ to $K = 0$ and back to $K > 0$, at $r = 1$.  We also show the same plots for symplectic Euler, Eq. \eqref{eq:simpsplit}, or, equivalently, Eq. \eqref{eq:split1} with $K = 0$, in Fig. \ref{fig:euler}.  There are $10$ small symplecticity error spikes with different origin: these occur when the particle reaches periapse.  

\begin{figure}
    \centering
    \resizebox{0.99\columnwidth}{!}{\includegraphics{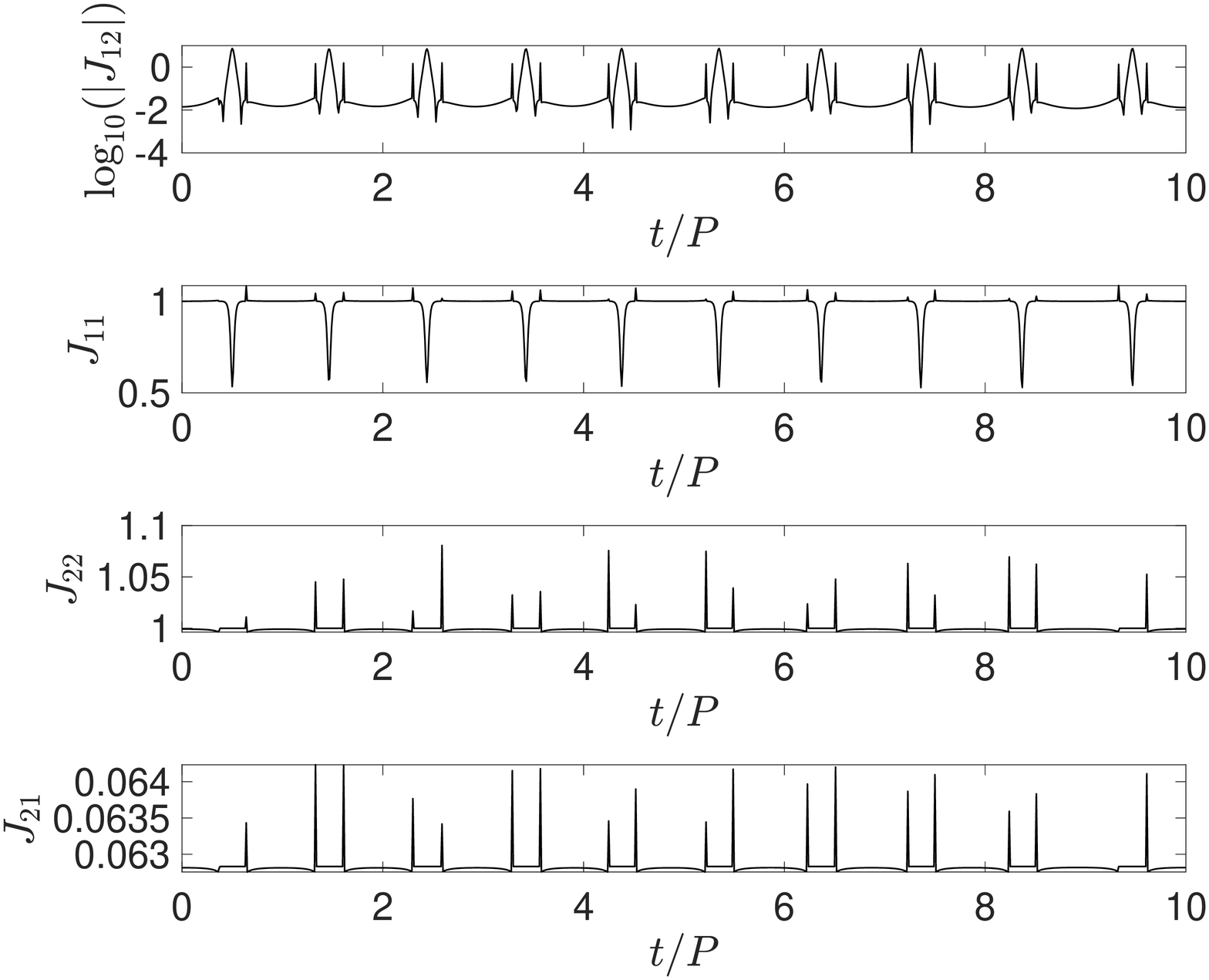}}
    \caption{Jacobian matrix elements for a hybrid integrator with transition function \ref{item:lin}.  The definition of the Jacobian matrix elements is given by Eq. \eqref{eq:jacdef}.  The greatest variation is seen in the top two panels, which measure variations in $p$.  A $\log$ scale is used to observe large variations in $J_{12}$.  The ruggedness of $J_{1 2}$ with time explains where the symplecticity errors for this transition function come from.}
    \label{fig:d0f2}
\end{figure}

\begin{figure}
    \centering
    \resizebox{0.99\columnwidth}{!}{\includegraphics{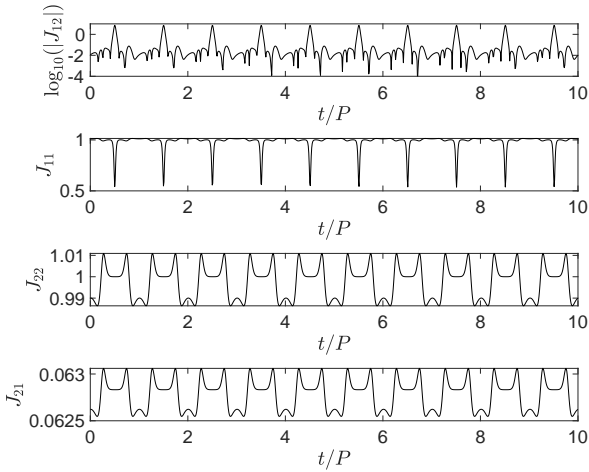}}
    \caption{{Same as in Fig. \ref{fig:d0f1}, but for an integration with transition function \ref{item:smooth}.}  $J_{1 2}$ changes more smoothly as compared to the transition function \ref{item:lin} plots.  Symplecticity errors are observed due to finite differencing errors making $J_{1 2}$ look rougher than it actually is.}
    \label{fig:dallf2}
\end{figure}

Does the appearance of the symplecticity spikes in all integrators mean our prediction of which functions are symplectic was wrong?  The answer is no: our measure of symplecticity is imperfect in a few ways.  To explore this issue further, let us take a look at the Jacobian elements of the hybrid integrators with functions \ref{item:lin} and \ref{item:smooth} in Fig's. \ref{fig:d0f2} and \ref{fig:dallf2}, respectively.    

The Jacobian elements with the greatest variation are in the top two panels, $J_{1 1}$ and $J_{1 2}$.  This is explained because they measure variations of $p$, which depend on the potentially rapidly evolving $K$ and $dK/dr$.  These two matrix elements have spikes at the same times of the symplecticity spikes.  The variations in $J_{1 2}$ are greater than the other matrix elements, so we use a $\log$ scale to view its variations.  If we look closely, we note that the topology of $J_{1 2}$ as a function of time is rugged in Fig. \ref{fig:d0f2} while it is smoother in Fig. \ref{fig:dallf2}.  The ruggedness of $J_{1 2}$ for the linear function explains where its symplecticity errors come from.  $J_{1 2}$ changes more gradually for the $\tanh$ function, but finite differencing approximations to derivatives make $J_{1 2}$ appear rougher than it is, leading to unphysical symplecticity errors.  Note the integrator proceeds normally and is symplectic or not regardless of the accuracy of our Jacobian estimates.    

We look more closely at some of the numerical issues we encounter in these tests.  For case \ref{item:lin}, at the $38$th step, the Bulirsch--Stoer integrator uses values of both $dK/dr = 0$ and $dK/dr = 1$ in its polynomial extrapolation for the estimate of the state.  The rapid change in the derivative breaks symplecticity and leads to energy drift.  Bulirsch--Stoer is not as suitable as other high-accuracy integrators for evaluating non-smooth functions.  We repeated this test by substituting Bulirsch--Stoer for an adaptive step fourth and fifth order Runge--Kutta--Fehlberg (RKF) method, which is more suitable for navigating the terrain of non-smooth functions \citep{press02}.  Our conclusions are the same when we use this RKF method instead.  An error tolerance of $10^{-12}$ was used for both Bulirsch--Stoer and the RKF method.

We explore this 38th step further.  {The following refers to methods for solving this one step.}  Rather than use a Bulirsch--Stoer or RKF method, we take $100$ small symplectic Euler steps that solve Eq. \eqref{eq:compmap} alone.  We verify the method of taking small Euler steps converges by doing the following: we run $2^n$ {(instead of 100)} symplectic Euler steps {to solve \eqref{eq:compmap} only.  $H_2$ is the first Hamiltonian to get solved, so the initial conditions are those at the start of step 38}.  $n$ is a positive integer.  The size of each Euler step is $h_{\mathrm{Euler}} = P/(100\times 2^n)$.  We obtain solution $(r_1,p_1)$.  Then we run $2^{n+1}$ steps of stepsize $P/(100\times 2^{n+1})$, yielding $(r_2,p_2)$.  A quantity {$e = \sqrt{(r_2 - r_1)^2 + (p_2 - p_1)^2}/(h/2)$} is calculated.  If the method converges, we expect $e$ to scale linearly with $h_{\mathrm{Euler}}$.  Note that $h$ is a constant, independent of $n$.  Indeed, we calculate a slope of $1.15$ on a $\log e$--$\log h_{\mathrm{Euler}}$ plot using the points $n = 1$ to $n = 11$. but with a weak correlation $R^2 = 0.68$.  With a twice differentiable transition function (not described in Section \ref{sec:sfunc}), the slope is $1.061$ with better correlation $R^2 = 0.99$.  The small Euler step method still converges despite the discontinuities present.  We made sure not to use $n$ too large such that roundoff error would affect the power law calculation: the difference $|r_2 - r_1|$ was always greater than $10^{-10}$.  By using leapfrog steps instead of symplectic Euler steps, we again verify linear convergence when discontinuities are present (convergence is quadratic with the twice differentiable smoothing function).  Having verified the 100 Euler step method converges, we rerun the Kepler test, substituting this method for Bulirsch--Stoer.  The linear transition function still yields unstable energy error, while the twice differentiable function yields stable energy error.  Thus, we checked through various methods that the method for solving Eq. \eqref{eq:compmap} does not affect our conclusions.

When an integrator is non-symplectic, there is a sudden and rapid change in the equations of motion, leading to a rapid dynamical timescale the integrator cannot resolve.  Even if a transition function is symplectic, if it transitions too rapidly, the timescale will also not be resolved by the integrator and the energy error will grow secularly.  In the linear transition function test, there is a jump in the first term of \eqref{eq:compmap2}, while the second term is approximately 0.  For a case where the first term is approximately $0$, while the second jumps, we use function \ref{item:smooth} with $k = 100$.  In this case, the energy error also grows secularly.  In fact, it grows for $k \gtrsim 35$.  At $k \approx 35$, \eqref{eq:compmap2} gives $\dot{p} \approx 20$ at $x = 0$, so that $p$ jumps about $7\%$ of its total range during the crossing timestep.  During the jump, the denominators in \eqref{eq:compmap2} are $\approx 1$.  Finally, we can vary the size of the jump in this test by using $k = 100$ and changing the $1/2$ to a constant $c$ in \ref{item:smooth}.  One might suspect if the jump discontinuity is small enough, no secular drift will occur because the numerical method cannot detect the non-smoothness.  Indeed, a secular drift is detected only if $c \gtrsim 0.025$.  Thus, violations in the Lipschitz continuity of the ODEs were allowed if the jump discontinuities are small enough.

The Heaviside \ref{item:heavi} and modified Heaviside \ref{item:modheavi} functions lead to energy drift and symplecticity error jumps as expected.  For the modified Heaviside function, the integrator transitions from $K = 0$ to $K = 1$ at the following steps with irregular spacing: 82, 180, 276, 370, 463, 554, 644, 733, 821, 909, 997.  For the integrator to be symplectic, these steps would need to be regularly spaced.  $dK/dr$ is always $0$ for this integrator.  

{Another question is whether a problem point is actually hit.  For integrations with transition function \ref{item:heavi}, we verified this situation did not occur in our tests.  To get a sense of the probabilities for landing on a potentially problematic point, consider an $e = 0$ orbit.  The circle is described by approximately $p = 2 \pi \times 10^{16}$ double precision numbers.  The chance of randomly picking a particular point is $p^{-1}$.   Landing or not on a point with discontinuities is not a main concern in this work.}

There are other causes of energy drift for symplectic integrators related to rapid timescales.  For example, symplectic Euler gives a secular error increase when solving a Kepler problem that is too eccentric.  {For a given eccentricity, a minimum timestep resolves periapse} (see also, \cite{W15} for requirements so that an integrator resolves periapse).  Indeed, rapid timescales are the reason close encounters are so problematic in $N$-body problems in the first place, they occur on timescales that are too rapid for the integrator to resolve.  A rigorous analysis of what timescales are problematic for an integrator of given stepsize is beyond the scope of this work, but clues on how to approach this problem are given in \cite[Section 2.3]{leim04}.

\subsection{Long term energy drift}
\begin{figure}
    \centering
    \resizebox{0.99\columnwidth}{!}{\includegraphics{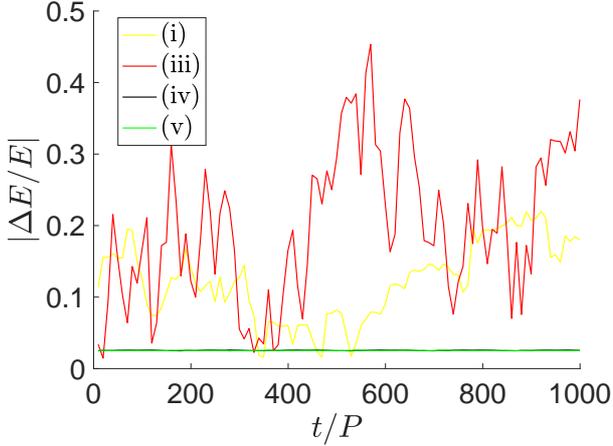}}
    \caption{Long term energy error evolution for hybrid integrators with different switching functions.  We integrate a Kepler problem with $e = 0.7$, $h = P/100$, and $t = 1000 P$.  The initial conditions are at apoapse: $r = 1+e$ and $p = 0$.  The median absolute energy error every 100 steps is plotted.  The symplectic everywhere integrators show stable energy error while the integrators that are symplectic everywhere except at two points in phase space (the integrator differential equations are integrable) show erratic energy error behavior.}
    \label{fig:longterm}
\end{figure}

We do a long term energy error test with selected non-symplectic and symplectic transition functions.  We use the same eccentricity and timestep, run for $t = 1000 P$, and calculate the energy error at each step.  To avoid large variations in energy errors, we plot the median absolute energy error every 100 steps (so about one point per period).  {The conclusions of this test are unchanged by instead studying the mean.}  The result is shown in Fig. \ref{fig:longterm}.

It's seen the integrators that are non-symplectic at the two points have an erratic error behavior while the fully symplectic integrators {have no clear energy drift.  No clear energy drift is seen even after zooming in vertically by over a factor of 100.}

One option we can consider is whether the drifts are caused by $h$ being too large, leading to stepsize chaos.  If this is true, we should be able to reduce $h$ and find one of the irregular curves of Fig. \ref{fig:longterm} becomes regular.  We repeated the linear transition function integration with step $100$ times smaller, and still found no regularity.  According to Section \ref{sec:sympint}, without Lipschitz continuity in the ODEs, notion of periodic orbits are undefined.  We ran the linear function integration for a longer time and found that by $t = 10000 P$, the orbit had become hyperbolic, which is unphysical.  {Symplectic integrators are used over long dynamical timescales, such as the Solar system age.  These results indicate that over only 10,000 periods, which equates to 10,000 years for Earth's orbit, or $< 10^{-5}$ the Solar system age, certain hybrid integrators give unphysical solutions.}

We also tested whether our results hold for more degrees of freedom.  We considered the two degree of freedom Hamiltonian from \eqref{eq:kep2}.  A hybrid integrator is constructed using,
\begin{equation}
H_1 = \frac{p_1^2+p_2^2}{2} - \frac{K(r)}{\sqrt{x_1^2 + x_2^2}} \quad \text{and} \quad H_2 =  - \frac{1}{\sqrt{x_1^2 + x_2^2}}\left(1 - K(r) \right).
\end{equation}
Now, solving each Hamiltonian requires solving a system of four first order ODEs.  Note $H_1$ and $H_2$ are integrable.  The initial conditions at apoapse in these coordinates are chosen to be $x_1 = 1+e$, $x_2 = 0$, $p_1 = 0$, and $p_2 = \sqrt{(1-e)/(1+e)}$.  After constructing this hybrid integrator, and leaving all other integration parameters the same, we found the linear transition function still yields energy drift, while function \ref{item:polyn} does not.  

\begin{figure}
    \centering
    \resizebox{0.99\columnwidth}{!}{\includegraphics{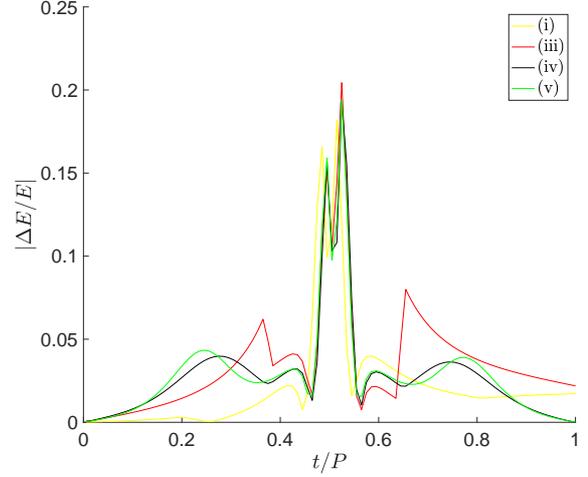}}
    \caption{Energy error over one period for the four curves of Fig. \ref{fig:longterm}, but the error is plotted at each timestep.  The symplectic curves are symmetric in the region $K > 0$, leading to long-term energy conservation.}
    \label{fig:oneenc}
\end{figure}

Over one period, the difference between the four curves of Fig. \ref{fig:longterm} is arguably less clear.  We plot the energy error for the four integrators each time step for one period in Fig. \ref{fig:oneenc}.  The symplectic curves show nearly symmetric error behavior for the region $K > 0$.  This symmetry is lost in the other curves.  The symmetry in the symplectic curves ensures long-term error conservation.
\ref{fig:oneenc}.

{
We wish to confirm our results with different Hamiltonians.  We test the simple harmonic oscillator, with Hamiltonian $H_{\text{sho}} = (q^2 + p^2)/2$.  The period is again $P = 2\pi$.  We can use the symplectic Euler method again, with splitting,}

\begin{equation}
H_1 = \frac{q^2 K(q) + p^2}{2} \quad \text{and} \quad H_2 =   \frac{{q^2} (1 - K(q))}{2}.
\end{equation}
{
For initial conditions, we choose $q = 1$ and $p = 0$, so that $H_{\text{sho}} = +1/2$.  $H_2$ is solved easily while, again, Bulirsch--Stoer is used to solve $H_1$.  $G(x)$ is the same as before, but now $x = (q+2)/2$.  In the exact solution, $x \in (1/2,3/2)$.  In our experiments, $x$ will be able to take on negative values and values $> 1$ each period.  In order, the four discontinuities we integrate over each period are encountered in the transitions,
\begin{itemize}
\item $K = 1$ to $0 < K< 1$
\item $0 < K < 1$ to $K = 0$
\item $K = 0$ to $0 < K < 1$
\item $0 < K < 1$ to $K = 1$.
\end{itemize}
The stepsize is chosen as $h = P/100$ and the runtime is $t = 10,000 P$.  We measure deviations in the energy from $1/2$.  The conclusions we tested from the previous experiments stayed the same.  In particular, periodic energy errors are found using transition function \ref{item:polyn}, while energy drift is observed with function \ref{item:lin}.}

\section{Conclusion}
\label{sec:conc}
This paper seeks to test the assumption that symplectic integrators can safely break symplecticity if the breaks are only at a few points in phase space.  This assumption is made in a wide variety of $N$-body codes at all scales; for example in hybrid or multiple time-stepping codes.  We considered a hybrid symplectic integrator which is symplectic everywhere or breaks symplecticity at two points in phase space.  The hybrid integrators with symplecticity breaks were significantly less stable.  They did not necessarily hit the problem points, but they integrated over them.  Breaking symplecticity introduces unresolved timescales in the integrators.  Physical or numerical mechanisms, such as scattering of planets, also can introduce such unresolved timescales and lead to deterioration in the integrator performance.  We showed how to correct this deterioration in the case of hybrid symplectic integrators, by ensuring the Lipschitz continuity of the equations of motion.  Lipschitz continuity is only required over the domain of the $N$-body map: if a discontinuity exists but an $N$-body method does not integrate over it, there is no problem.  In the case of multiple time stepping cosmological schemes \citep{S05} or block time-step schemes \citep{farr07}, this paper does not offer a solution to this problem{, nor does it study how serious the problem is.}  {It is worth mentioning the Hamiltonian splits in this paper involved splitting an integrable Hamiltonian into two integrable pieces.  It is conceivable to split an integrable Hamiltonian into nonintegrable pieces, but we could not concoct a practical situation in which this would be useful.  So we have not tested this scenario, which could hypothetically change some result.}  When possible, fully symplectic integrators should be utilized to solve the $N$-body problem.

\section{Acknowledgements}
I appreciate discussions with Hanno Rein, Dan Tamayo, Scott Tremaine, Ed Bertschinger, {Walter Dehnen}, Matt Payne, and Marie--Claude Arnaud.  {I appreciate feedback from the anonymous referee.}

\appendix
\section{Comparison with previous result}
\cite{H16} found that the hybrid code \texttt{MERCURY} \citep{C99} was non symplectic, a result explained in other work \citep{Reinetal2019}.  \cite{H16} stated that unless a hybrid integrator is $C^\infty$ over some domain, its perturbed Hamiltonian \citep{hair06} is undefined, and it was concluded the integrator cannot be exactly symplectic.  How we define symplecticity is clearly important.  We discussed in Section \ref{sec:sympint} that a notion of symplecticity can exist even for $C^0$ Hamiltonians.  We have decided in this paper to define symplectic integrators as those derived from at least $C^{1,1}$ Hamiltonians.  According to this definition, hybrid methods can be symplectic.  The concept of symplecticity is an active area of research \citep{BHS18}.  Numerically, we have found in this work even a symplectic Euler method shows breaks in symplecticity, but argued this is due to our limitation in measuring Jacobians.

\bibliographystyle{mnras}
\bibliography{refs}
\end{document}